\documentclass[reprint,amsmath,amssymb, aps, prl,
floatfix,nofootinbib]{revtex4-1} 
\usepackage{graphicx}
\usepackage{dcolumn}
\usepackage{bm}
\usepackage{caption}
\usepackage{amssymb,amsmath,color}
\usepackage{gensymb}
\usepackage{url}
\usepackage{float}
\usepackage{braket}
\usepackage{hyperref}    
\hypersetup{colorlinks = true, linkcolor = blue, anchorcolor = blue, citecolor = blue, filecolor = blue, urlcolor = blue}
\usepackage{xcolor}
\usepackage{float}
\usepackage{tikz}
\usepackage{pgfplots}

\begin{document}

\title{Electrons surf phason waves in moir\'{e} bilayers}

\author{Indrajit Maity} 
\author{Arash A. Mostofi}
\email{a.mostofi@imperial.ac.uk}
\author{ Johannes C. Lischner} 
\email{j.lischner@imperial.ac.uk}
\affiliation{Departments of Materials and Physics and the Thomas
Young Centre for Theory and Simulation of Materials, Imperial
College London, South Kensington Campus, London SW7 2AZ, UK}


\begin{abstract}
We investigate the effect of thermal fluctuations on the atomic and electronic structure of a twisted MoSe$_{2}$/WSe$_{2}$ heterobilayer using a combination of classical molecular dynamics and \textit{ab-initio} density functional theory calculations. Our calculations reveal that thermally excited phason modes give rise to an almost rigid motion of the moir\'e lattice. Electrons and holes in low-energy states are localized in specific stacking regions of the moir\'e unit cell and follow the thermal motion of these regions. In other words, charge carriers surf phason waves that are excited at finite temperatures. Small displacements at the atomic scale are amplified at the moir\'{e} scale, which gives rise to significant surfing speeds. We also show that such surfing survives in the presence of a substrate and disorder. This effect has potential implications for the design of charge and exciton transport devices based on moir\'{e} materials.

\end{abstract}

\maketitle

\textit{Introduction.} 
Moir\'e  materials, in which two or more two-dimensional (2D) materials are stacked and rotated relative to one another, have emerged as a new platform to discover, understand and manipulate electronic properties, such as superconductivity and correlated insulating states~\cite{Andrei2021the, Andrei2020graphene}. Twisting causes the electronic states to localize in specific regions of the moir\'e lattice, which results in a flattening of the corresponding bands and the emergence of strong electronic correlations. Because of this, moir\'e materials have been proposed as simulators of novel phases in quantum condensed matter~\cite{Kennesmoire2021}.

The effect of the twist on the electrons is often described in terms of an effective moir\'e potential which traps the electrons in specific regions of the moir\'e lattice. Recent experiments on twisted bilayer transition metal dichalcogenides have estimated that the variation of this moir\'{e} potential across the moir\'e unit cell can be as large as 300~meV~\cite{Shabanideep2021}. 

To date, most theoretical studies have investigated the electronic structure of twisted multilayer systems at zero temperature and, hence, in a static moir\'e potential~\cite{Naikultraflat2018, Mattiagamma2021, Xianrealization2021, Shabanideep2021, Indrajitreconstruction2021}. However, temperature effects can be significant in moir\'e materials. This is because of so-called \emph{moir\'e amplification}, whereby small atomic-scale thermal motion can give rise to large displacements of the moir\'e sites.~\cite{Liantwisted2019} At finite temperatures, therefore, the moir\'e potential is dynamic and the charge carriers reside in a highly mobile trapping potential. The behaviour of a quantum particle in such a time-dependent trapping potential is an interesting problem~\cite{Doescherinfinite1969, Miguelquantum2020, Granotquantum2009} with applications to quantum transients~\cite{Delcampoquantum2009}, transport of small particles using scanning tunnelling microscopy or optical tweezers, and charge transport~\cite{Thoulessquantization1983}. 

In this Letter, we study the behaviour of localized electrons and holes in twisted MoSe$_{2}$/WSe$_{2}$ heterobilayers at finite temperatures using a combination of classical molecular dynamics simulations and \textit{ab initio} density-functional theory calculations. We find that the moir\'{e} sites are highly dynamic, which is a result of very soft phason modes that are thermally excited. Charge carriers follow the motion of the moir\'{e} sites to which they are localized, and behave as though they are surfing on the dynamic moir\'{e} potential. The surfing speed is significantly different for twist angles close to 0$^\circ$ as compared to those close to $60^\circ$. We also discuss the impact of static disorder and a substrate on the surfing charge carriers. 

\textit{Methods.} Atomic structures of flat twisted MoSe$_{2}$/WSe$_{2}$ heterobilayers were generated using the TWISTER package~\cite{Naiktwister2022}. Because of the large size of the moir\'{e} unit cell, we used classical interatomic potentials fitted to \textit{ab initio} density functional theory calculations to determine the relaxed atomic positions, study phonon properties and carry out molecular dynamics simulations. Specifically, interactions between atoms in the same layer, i.e., intralayer interactions, were described using a Stillinger-Weber potential~\cite{Zhouhandbook2017}, while interlayer interactions were described using a Kolmogorov-Crespi potential~\cite{Naikkolmogorov2019}. Structural relaxations and molecular dynamics simulations were performed using the LAMMPS package~\cite{Thompsonlammps2022,lammps} and phonon calculations were performed using a modified version of the PHONOPY package~\cite{Togofirst2015}. For the molecular dynamics simulations, a $3\times3$ moir\'{e} supercell was used. Larger supercells (up to $20\times20$) were used to verify the system-size convergence of our results.

Electronic structure calculations using \textit{ab initio} density functional theory as implemented in the SIESTA package~\cite{Solersiesta2002} were performed on atomic structures obtained from the classical potentials. We included spin-orbit coupling~\cite{Seivaneon2006} in all our calculations. We used norm-conserving Troullier-Martins pseudopotentials~\cite{Troullierefficient1991} and the local density approximation to describe exchange-correlation effects~\cite{Perdewself1981}. Further details are provided in Sec.~I of the Supplementary Information (SI)~\cite{SI}. 

\begin{figure}[!ht]
\centering
\includegraphics[scale=0.4]{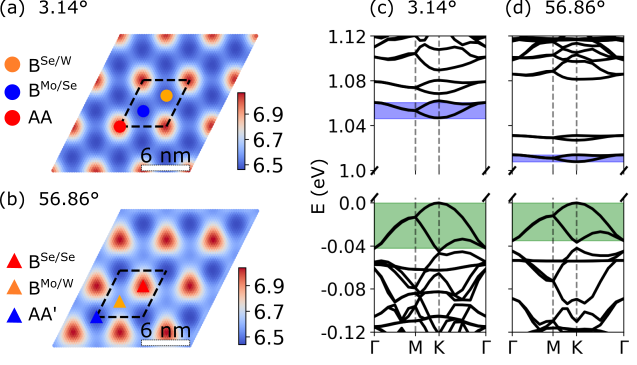}
\caption{Interlayer separation (ILS) landscape for relaxed 3.14$^\circ$ (a) and $56.86^\circ$ (b) twisted MoSe$_{2}$/WSe$_2$ heterobilayers. The colour bar is in units of angstrom. The positions of high-symmetry stackings in the moiré unit cell (dashed line) are indicated by symbols. (c),(d):  Corresponding electronic band structures. The widths of the highest valence band and lowest conduction band are indicated by shaded areas. The zero of energy is set to the valence band maximum.}
\label{fig1}
\end{figure}

\textit{Structural and electronic properties of twisted MoSe$_2$/WSe$_{2}$ at zero temperature.} 
In Fig.~\ref{fig1}(a) and (b) we show the interlayer separation (ILS) landscape for relaxed twisted MoSe$_2$/WSe$_{2}$ bilayers with twist angles of 3.14$^\circ$ and 56.86$^\circ$, respectively. For twist angles close to $0^\circ$, the moiré unit cell contains three high-symmetry stackings, which we label AA (Mo above W and Se above Se), $\mathrm{B^{Mo/Se}}$ (Bernal stacking with Mo above Se), and $\mathrm{B^{Se/W}}$ (Bernal stacking with Se above W). Among these, $\mathrm{B^{Mo/Se}}$ is energetically the most favourable. 
On the other hand, for twist angles close to $60^\circ$, the high-symmetry stackings in the moiré cell are AA$^\prime$ (Mo above Se and Se above W), $\mathrm{B^{Mo/W}}$ (Bernal stacking with Mo above W), and $\mathrm{B^{Se/Se}}$ (Bernal stacking with Se above Se). Among these, $\mathrm{AA^\prime}$ is energetically the most favourable. 
In agreement with recent experiments~\cite{Shabanideep2021}, the calculated ILS landscape has a six-fold symmetry around the AA stacking site for systems with twist angles near 0$^\circ$ and a three-fold symmetry around the $\mathrm{B^{Se/Se}}$ site for systems with twist angles near 60$^\circ$.

Figures~1(c),(d) show the electronic band structures of relaxed 3.14$^\circ$ and $56.86^\circ$ twisted heterobilayers, respectively. These systems form type-II heterostructures with the valence band maximum (VBM) derived from the K valley of the WSe$_{2}$ layer and the conduction band minimum (CBM) from the K valley of the MoSe$_{2}$ layer. The widths of the VB and CB are smaller for twist angles close to $60^\circ$. For example, at 3.14$^\circ$, the VB width is 42 meV and the CB width is 14 meV, while at 56.86$^\circ$ the VB with is 35 meV and the CB width is only 6 meV. To understand the origin of these differences, we also perform calculations on individual layers with the same atomic structure as in the relaxed twisted system (see Sec.~II of the SI~\cite{SI} for details). For the 56.86$^\circ$ system, the highest valence and the lowest conduction band from the corresponding monolayer calculations agree well with the twisted bilayer result. In contrast, this is not the case for the 3.14$^\circ$ system indicating that interlayer tunnelling plays a more important role for twist angles near 0$^\circ$ than for twist angles near 60$^\circ$. Similar results have been reported for twisted homobilayers of transition metal dichalcogenides~\cite{Fengchenghubbard2018, Kundumoire2022}. Homobilayers with twist angles close to $60^\circ$ exhibit an inversion symmetry which - in combination with time reversal symmetry - enforces the degeneracy of K-derived bands indicating that off-diagonal interlayer tunnelling terms vanish. In the corresponding heterobilayers, this inversion symmetry is broken because the metal atoms in both layers are different, but despite this, the interlayer tunneling is much weaker than in the systems with twist angles near $0^\circ$.

\begin{figure*}[!ht]
\centering
\includegraphics[scale=0.7]{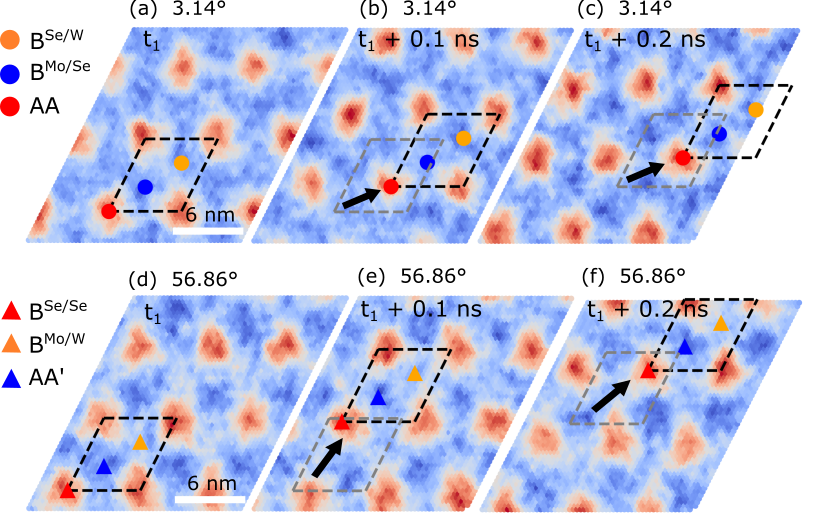}
\caption{
Time evolution of the interlayer separation landscape of twisted MoSe$_2$/WSe$_2$ heterobilayers during molecular dynamics simulations at $T=150$~K, for twist angles of 3.14$^\circ$ [(a)-(c)] and $56.86^\circ$ [(d)-(f)], with snapshots separated by 0.1~ns. The colour bar is in units of angstrom. The positions of the high-symmetry stackings within the moir\'e unit cell are marked by symbols. The direction in which the moir\'{e} lattice moves is indicated by the black arrow and the unit cell of the previous snapshot is marked with grey colour.}
\label{fig2}
\end{figure*}

\begin{figure*}[!ht]
\centering
\includegraphics[scale=0.25]{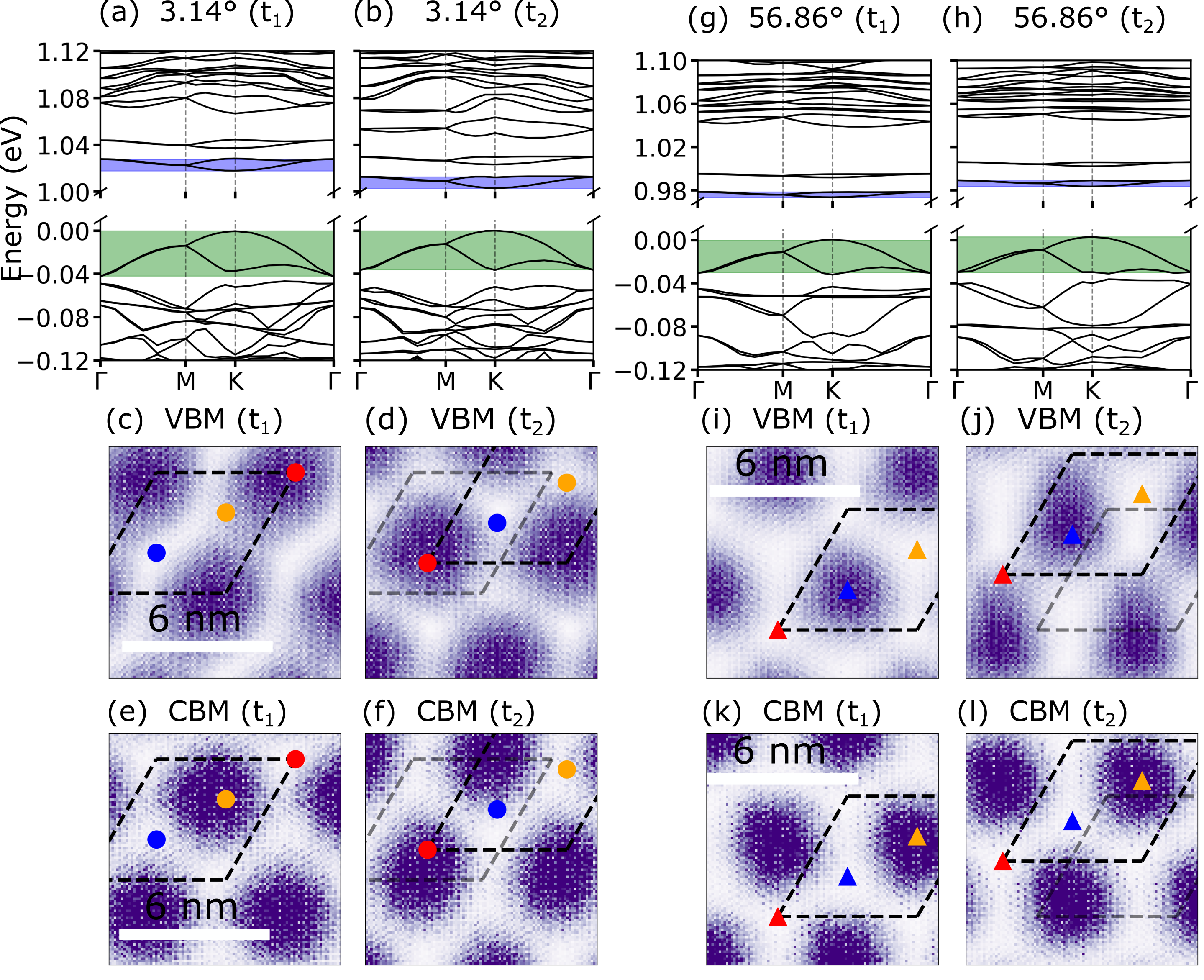}
\caption{Top panels ((a)-(b),(g)-(h)): electronic band structures for atomic structures obtained from molecular dynamics simulations at $T=150$~K for a twisted MoSe$_2$/WSe$_2$ heterobilayer at two different times ($t_{1}$ and $t_{2}=t_{1}+0.2$~ns). Bottom panels ((c)-(f),(i)-(l)): the squared absolute magnitudes of the VBM and CBM wavefunctions at the $\Gamma$-point of the moir\'e unit cell for the same atomic structure as in the top panels. The wavefunctions are averaged over the out-of-plane direction. The different high-symmetry stacking regions in the moir\'e unit cell are indicated by symbols.}
\label{fig3}
\end{figure*}

\textit{Structural and electronic properties at finite temperatures.} Fig.~\ref{fig2} shows ILS landscapes of snapshots (separated by 0.1~ns) from the molecular dynamics simulations at $T=150$~K. It is observed that the moir\'e pattern stays largely intact, but moves as a whole by several nanometers. Similar results are also observed for smaller twist angles~\footnote{However, no such movement of the moir\'e lattice is observed for untwisted bilayers, i.e. at twist angles of $0^\circ$ and $60^\circ$.}.

We interpret the rigid motion of the moir\'e lattice at finite temperatures to be a consequence of thermally excited phason modes. A phason represents an effective global translation of the moir\'{e} lattice due to the uniform relative displacement of the two layers~\cite{Ochoamoire2019, Maityphonons2020}~\footnote{Note that the actual polarization vectors often display a more complicated spatial structure, see Sec. III of the SI~\cite{SI}.}. The energy cost associated with such modes is very small ($<$ 0.1 meV)~\cite{Ochoamoire2019, Maityphonons2020, Jonathanlow2022, Koshinomoire2019, Samajdarmoire2022, Gaosymmetry2022,krisnamoire2022}. Therefore, these modes are easily excited at finite temperatures.

Figures~\ref{fig3}(a), (b), (g) and (h) show the electronic band structures obtained for different molecular dynamics snapshots. The band structures are qualitatively very similar to those obtained for the relaxed atomic structure at $T=0$~K. However, the electronic band gap is renormalized due to thermal fluctuations. In particular, we observe a band-gap reduction of $37\pm10$~meV for 3.14$^\circ$ and $30\pm8$~meV for 56.86$^\circ$ systems at 150~K. Interestingly, we find that electrons and holes follow the movement of the moir\'e sites. For the 3.14$^\circ$ system, the VBM (CBM) always remains localized on the AA ($\mathrm{B^{Mo/Se}}$) site [see Figs.~\ref{fig3}(e)-(f)], while for the 56.86$^\circ$ system, the VBM remains localized on the $\mathrm{AA^\prime}$ site and the CBM on the $\mathrm{B^{Mo/W}}$ site [see Figs.~\ref{fig3}(i)-(l)]. In other words, the electrons surf the phason waves that are excited at finite temperatures.

We have also investigated the speed at which electrons surf. Interestingly, the moir\'e lattice amplifies atomic displacements and, therefore, small atomic displacements induced by thermal motion can give rise to significant displacements of the moir\'e sites and a large velocity of the surfing electrons. For instance, a displacement of 1~\AA\ in the $x$-direction of all atoms of the MoSe$_2$ layer in a 3.14$^\circ$ twisted heterobilayer gives rise to a displacement of the moir\'e sites by approximately 18~\AA\ in the $y$-direction. The ratio of the moir\'e displacement and the atomic displacement is approximately $\frac{a_{m}}{a} \approx \frac{1}{\theta}$, where $a_m$, $a$ and $\theta$ are the moir\'{e} lattice constant, the lattice constant of MoSe$_{2}$ and the twist angle, respectively. This is illustrated in SI, Sec.~IV. 

To analyze the speed of surfing quantitatively, we define the mean distance travelled by a specific moir\'{e} site (e.g. AA site) as 
\begin{equation}
d(t) = \left\langle \frac{1}{N}\sum_{i=1}^{N} \left|{\bf r}_{i}(t) - {\bf r}_{i}(0)\right| \right\rangle  = v_{m} t,
\end{equation}
where $N$ is the total number of moir\'{e} unit cells (i.e., the number of moir\'{e} sites) in the supercell, ${\bf r}_{i}(t)$ is the position of $i$-th the moir\'{e} site at time $t$, $ \langle\ \cdots \rangle$ denotes the average over different molecular dynamics trajectories and $v_{m}$ is the surfing speed. We used 50 separate molecular dynamics trajectories with each trajectory containing 21 snapshots spanning over 0.1 ns.  Interestingly, the electrons surf more slowly in systems with twist angles near $0^\circ$ than in systems with twist angles near $60^\circ$. For example, we find $v_m \approx 25$~m/s for $3.14^\circ$ and $v_m \approx 40$~m/s for $56.86^\circ$. 

It is interesting to compare the measured surfing speeds to the group velocity of phasons. Despite their low energy, phasons are gapped at the $\Gamma$-point and only exhibit a linear dispersion with a group velocity of $\gtrsim 10^2$~m/s at wavevectors not too close to $\Gamma$, see SI, Sec. VI for the phonon dispersion. Very close to $\Gamma$, the phason dispersion flattens corresponding to a reduced group velocity. This behaviour of the phason group velocity suggests that the observed surfing speeds are sensitive to the supercell size used in our simulations. Indeed, we find that the surfing speed is reduced in larger supercells, see Sec.~VII of the SI. However, finite surfing speeds are observed even for the largest simulation cells with supercell lengths of 0.1$\mu m$ and we expect that the surfing speed in experiments is determined by the sample size which typically ranges from a few $\mu m$ to a few hundred $\mu m$.

We also compute the mean-square displacement of the moir\'{e} sites 

\begin{equation}
D(t) = \left \langle \frac{1}{N}\sum_{i=1}^{N} |{\bf r}_{i}(t) -
{\bf r}_{i}(0)|^{2} \right  \rangle
\end{equation}

The mean-square displacement is found to be proportional to $t^{2}$
indicating a free propagation of moir\'e sites, see Sec.~V of the
SI~\cite{SI}.  In contrast, diffusive propagation would give rise
to a mean-square displacement proportional to $t$.

\begin{figure}[!ht]
\centering
\includegraphics[scale=0.65]{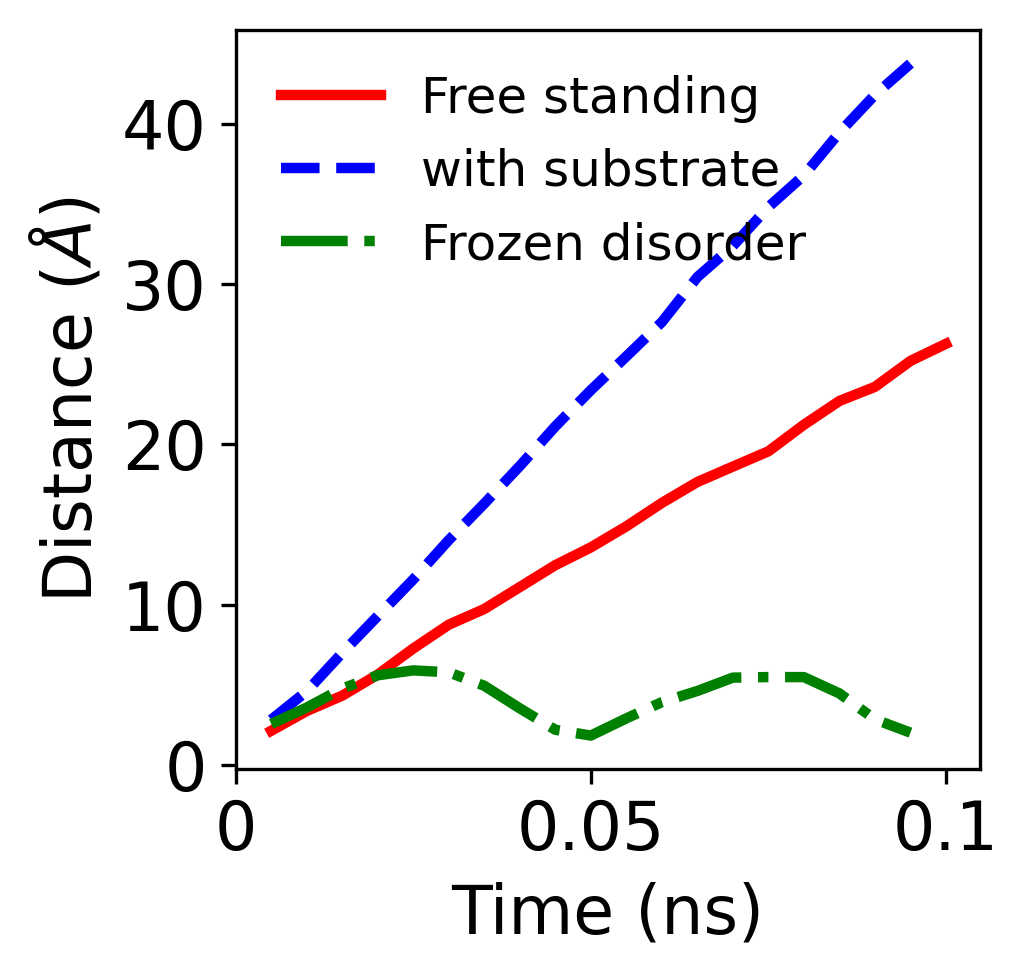}
\caption{Distance travelled by surfing charge carriers as function of time for (a) a free-standing twisted MoSe$_{2}$/WSe$_{2}$ bilayer, (b) a twisted MoSe$_{2}$/WSe$_{2}$ bilayer on a hexagonal boron nitride substrate, and (c) a twisted MoSe$_{2}$/WSe$_{2}$ bilayer in the presence of frozen disorder.} 
\label{fig4}
\end{figure}

\textit{Effect of substrate and disorder.} All results presented so far are for free-standing twisted heterobilayers. Experimental samples are typically placed on a substrate and contain disorder, for example, induced by extrinsic strain which can give rise to large twist-angle inhomogeneities~\cite{Shabanideep2021, Racheldirect2022}. To study the effect of a substrate, we include a hexagonal boron nitride (hBN) substrate in our molecular dynamics simulations. We also observe the motion of the moir\'e sites and charge carrier surfing in these simulations, as shown in Fig.~\ref{fig4} (blue, dashed line). Interestingly, the presence of a substrate increases the surfing speed: for a 3.14$^\circ$ twisted bilayer an increase of $\sim 40$ \% is found when a $3\times3\times1$ supercell is used.

To study the impact of static disorder, we first relax the twisted bilayer on the substrate and then perform a molecular dynamics simulation in which the substrate atoms are not allowed to move. Then, the substrate acts as a frozen disorder potential on the atoms in the twisted bilayer. In these simulations, we do not observe free propagation of moir\'e sites (Fig.~\ref{fig4}, green, dot-dashed line). This is a consequence of disorder-induced pinning of phasons. Such pinning was also found in twisted bilayer graphene in a recent theoretical study~\cite{Ochoadegradation2022}. Even though there is no free propagation of moir\'e sites, there are large displacements of $\sim 5$~\AA~of the sites around their equilibrium position resulting from the moir\'e amplification of thermal fluctuations.

The disorder-induced pinning can be overcome by increasing the temperature. In order to demonstrate this, we perform molecular dynamics simulations at $T=1200$~K in the presence of the disorder and observe again the free propagation of moir\'e sites (see SI, Sec. VIII). Significant movement of moir\'e sites due to finite temperature in twisted bilayer graphene was observed in a recent experiment~\cite{Dejongimaging2022}. Similarly, the free sliding of twisted bilayer graphene nanoflakes due to thermal fluctuations has been reported~\cite{Xiaofengsuperlubric2013}. This clearly indicates that high-quality twisted bilayer materials are key to observing the surfing motion.  


Another way to detect and exploit charge carrier surfing is through sliding one layer of the twisted heterobilayer slides relative to the other layer. This generates a motion of the moir\'e sites in the transverse direction, as shown in SI, Sec.~IV. As the electrons follow the motion of the moir\'e sites, a transverse current is generated. Specifically, if the MoSe$_{2}$ layer is pulled along the $x$-direction with a speed $v^{x}$, the moir\'{e} sites move along $y$-direction with a speed $v^{y}_{m} \approx \frac{v^{x}}{\theta}$. The current density that is generated by the electron surfing is given by $j = nev^{y}_{m}$, where $n$ is the charge density and $e$ is the electron charge. Electron transport in such a chiral moir\'e charge pump could be topological as discussed in a related context in twisted bilayer graphene using continuum models and tight-binding models~\cite{Fujimototopological2020, Zhangtopological2020, Yingtopological2020}.

\textit{Conclusion.} We have demonstrated that the moir\'e sites of a twisted MoSe$_{2}$/WSe$_{2}$ heterobilayer move at finite temperatures due to thermally excited phasons. Electrons and holes follow the motion of the moir\'e sites and thus surf the phason waves. We have also discussed the impact of the substrate and disorder on free surfing. Our findings are relevant for the design of transport devices based on moir\'e materials~\cite{Rossi2023phason}. 

\textit{Acknowledgements.} This project received funding from the European Union’s Horizon 2020 research and innovation program under the Marie Skłodowska-Curie Grant agreement No. 101028468. The authors acknowledge support from the Thomas Young Centre under Grant No. TYC-101. This work used the ARCHER2 UK National Supercomputing Service and resources provided by the Cambridge Service for Data-Driven Discovery (CSD3) operated by the University of Cambridge Research Computing Service, provided by Dell EMC and Intel using Tier-2 funding from the Engineering and Physical Sciences Research Council, and DiRAC funding from the Science and Technology Facilities Council, and Imperial College Research Computing Service, DOI: 10.14469/hpc/2232. 


%

\pagebreak
\widetext

\renewcommand{\thefigure}{S\arabic{figure}}
\renewcommand{\bibnumfmt}[1]{[S#1]}
\renewcommand{\citenumfont}[1]{S#1}

\section{I: Simulation details}

\subsection{Generation of structures}
All the structures for the following twist angles were generated by the TWISTER package~\cite{Naiktwister2022}. The unit-cell lattice constants for both the MoSe$_{2}$ and WSe$_{2}$ layer were set to 3.32 \AA~while generating the moir\'{e} patterns. 

\begin{table}[h!]
\centering 
\begin{tabular}{|m{3cm}|m{3cm}|m{3.5cm}|}
\hline
Twist angles & Number of atoms & Moir\'{e} length (in \AA) \\
\hline 
3.14$^\circ$ & 1986 & 60.4 \\
\hline  
56.86$^\circ$ & 1986 & 60.4 \\
\hline 
\end{tabular}
\caption{Moir\'{e} patterns of twisted bilayer of $\mathrm{MoSe_{2}/WSe_{2}}$ studied in this work.}
\end{table}

\subsection{Structural relaxations}
The moir\'{e} patterns are relaxed using the LAMMPS package with the Stillinger-Weber ~\cite{Zhouhandbook2017}, and Kolmogorov-Crespi \cite{Naikkolmogorov2019} potentials to capture the intralayer and interlayer interactions of the twisted bilayer of $\mathrm{WSe_{2}}$, respectively. The Kolmogorov-Crespi parameters used in this work can correctly reproduce the interlayer binding energy landscape, obtained using density functional theory. The atomic relaxations produced using these parameters are in excellent agreement with relaxations performed using density functional theory. We relax the atoms within a fixed simulation box with the force tolerance of $10^{-5}$ eV/\AA\ for any atom along any direction. 

\subsection{Molecular dynamics simulations}
We equilibrate the moir\'{e} material under periodic boundary conditions in the canonical ensemble at several temperatures for a nano-second using a Nose\'{e}-Hoover thermostat. We track the dynamics of moi\'{e} sites using the micro-canonical ensemble for several nanoseconds. To extract the speed of the moir\'{e} sites, we have used a $3\times3\times1$ moir\'{e} supercell. Additional molecular dynamics simulations were performed for smaller twist angles ($1^\circ$ and $59^\circ$). Similar movements of moir\'{e} sites were observed. 

\subsubsection{Inclusion of the substrate and disorder}
We use hexagonal Boron Nitride (hBN) as a substrate for the MoSe$_{2}$/WSe$_{2}$ twisted bilayer. The intralayer interactions were described using a Tersoff potential~\cite{Tersoffempirical1988} and the interlayer interactions were captured using a Kolmogorov-Crespi potential~\cite{Liimaging2021}. The lattice constant for hBN was 2.516 \AA~and $24\times 24$ unit cells were added to the twisted bilayer. As long as the dynamics of the hBN and the twisted bilayer are allowed at finite $T$, we find the moir\'{e} sites move. However, if we freeze the movements of hBN in all directions, the moir\'{e} sites stop free movement but show large displacements around a mean position. The frozen hBN atoms in such a scenario act as a frozen disorder potential which pin the phason motion.

\subsection{Electronic structure calculations}
We use a double-$\zeta$ plus polarization basis for the expansion of wavefunctions. For all the electronic structure calculations we use the $\Gamma$ point in the moir\'{e} Brillouin zone to obtain the converged ground state charge density. A large vacuum spacing of $20$ \AA\ is used in the out-of-plane direction for all the density functional theory (DFT) calculations. All the electronic structure calculations at finite temperature are performed using a snapshot of the moir\'{e} unit cell from the classical molecular dynamics simulations. For computing the averaged electronic band gap at 150~K, we used 6 snapshots from our molecular dynamics simulations. Note that all the electronic structure calculations were performed on the moir\'{e} unit cell (i.e., $1\times1\times1$ moir\'{e} supercell).  

\clearpage
\newpage

\section{II : Electronic band-structure of $3.14^\circ$ and $56.86^\circ$ twisted MoSe$_{2}$/WSe$_{2}$}
\begin{figure}[h!]
\centering
\includegraphics[width=0.6\textwidth]{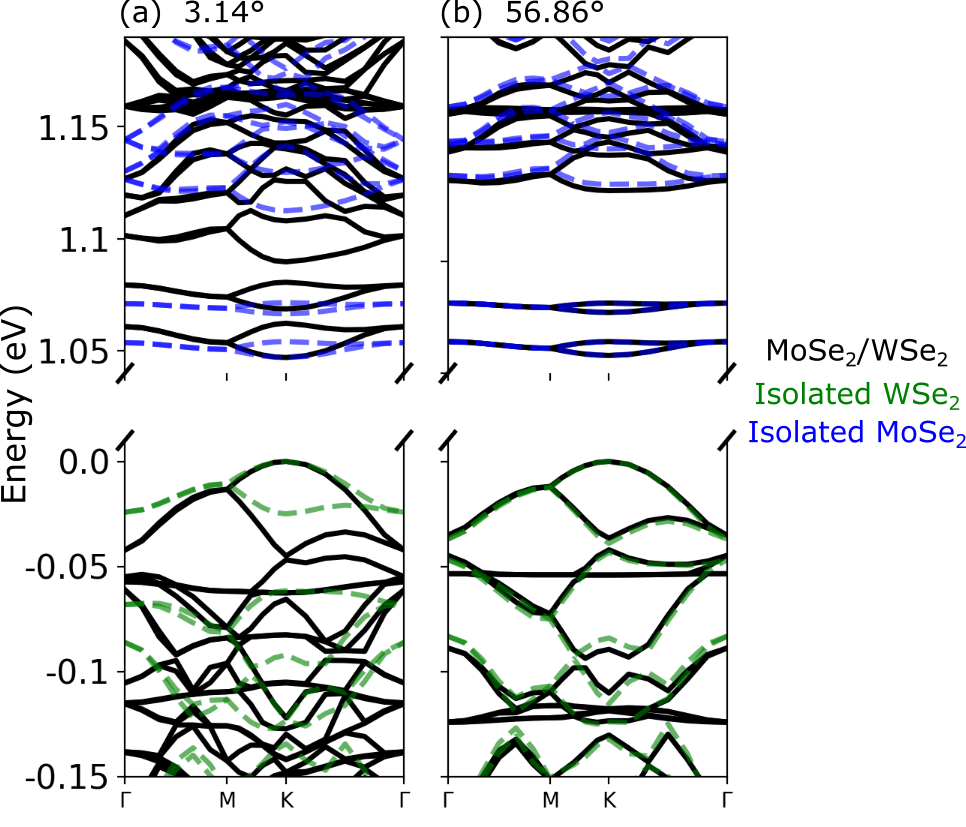}
\caption{Electronic band-structure calculations of 3.14$^\circ$ and $56.86^\circ$ twisted MoSe$_{2}$/WSe$_{2}$ heterobilayer. The band-structure calculations are compared with monolayer MoSe$_{2}$ and WSe$_{2}$ after isolating from the relaxed twisted bilayer calculations. No further relaxations on the isolated monolayers are performed. The valence band maximum of the twisted bilayer has been set to 0 eV. For better comparison, the valence band maximum of WSe$_{2}$ and the conduction band minimum of the MoSe$_{2}$ are aligned to those of twisted heterobilayer calculations.}  
\end{figure}

\clearpage 
\newpage 

\section{III : Phasons of $3.14^\circ$ twisted MoSe$_{2}$/WSe$_{2}$}
\begin{figure}[h!]
\centering
\includegraphics[width=1\textwidth]{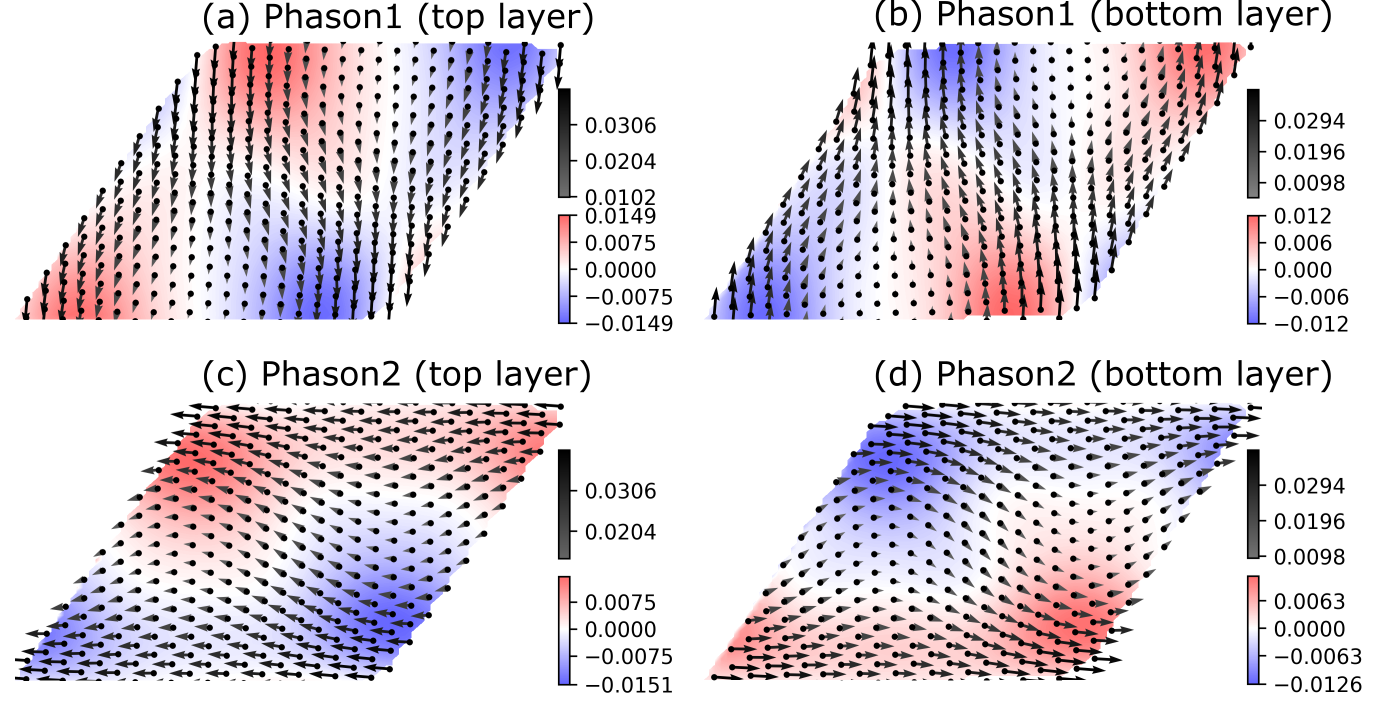}
\caption{Polarization vectors associated with the ultra-low frequency phason modes with energy 0.07 meV at the $\Gamma$ point shown for a moir\'{e} unit cell of the twisted MoSe$_{2}$/WSe$_{2}$ heterobilayer. The arrows (gray colorbar) denote in-plane displacements (only for Mo atoms of the top layer and W atoms for the bottom layer), whereas out-of-plane displacements are represented as a continuous field (colored). The origin of the arrow indicates the position of a metal atom.}  
\end{figure}

\clearpage
\newpage 

\section{IV: Illustration of moir\'{e} amplification}
\begin{figure}[!ht]
\centering
\includegraphics[scale=0.65]{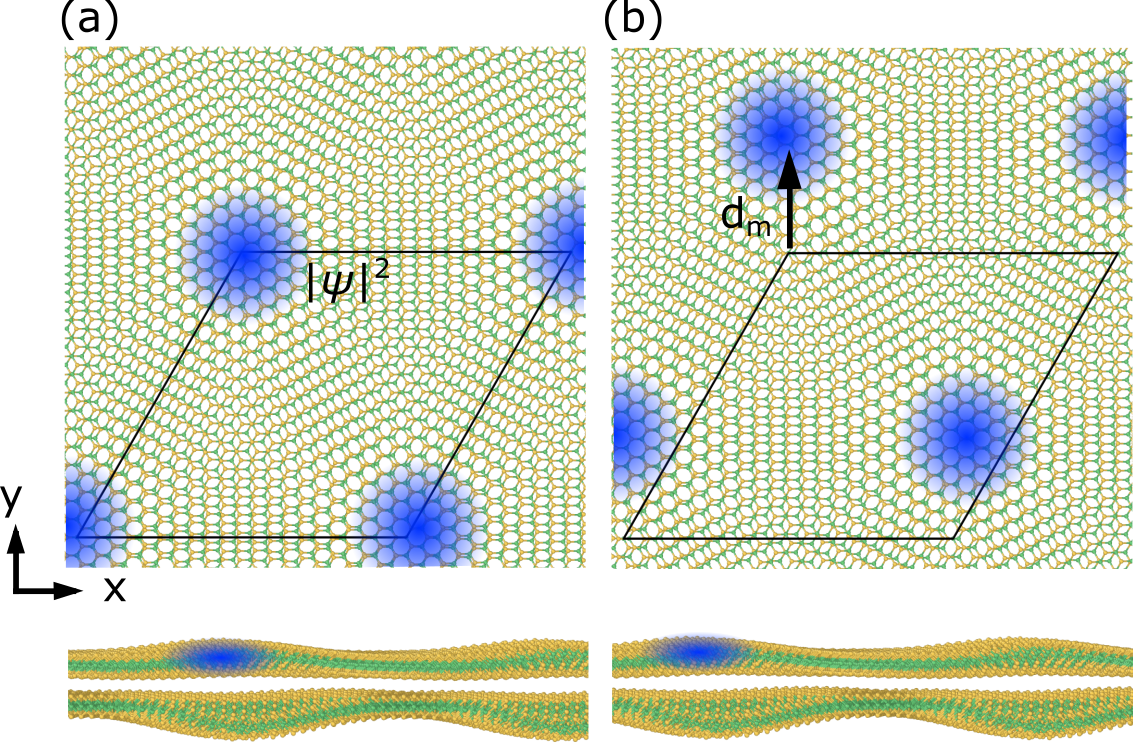}
\caption{Schematic of the moir\'e amplification of atomic displacements and the associated transport or ``surfing'' of charge carriers in a 3.14$^\circ$ twisted MoSe$_{2}$/WSe$_2$ bilayer. (a): Initial atomic structure and squared magnitude $|\psi|^{2}$ of the VBM wavefunction at the $\Gamma$-point of the moir\'e Brillouin zone. The moir\'e unit cell is indicated. (b) Atomic structure after the atoms in the top layer have been displaced by 1~\AA\ along $x$-direction. The AA site where the VBM is localized moves by $d_{m}\approx 18$ \AA\ along the $y$-direction. The top panels show the view from the top and the bottom panels show the view from the side. A moir\'{e} unit cell is indicated at a fixed position in both figures (solid black line). The out-of-plane displacements of the atoms are exaggerated for visual clarity in the bottom panels.} 
\label{figS3}
\end{figure}

\clearpage 
\newpage 
\section{V : Motion of Moir\'{e} sites}
\begin{figure}[h!]
\centering
\includegraphics[width=0.8\textwidth]{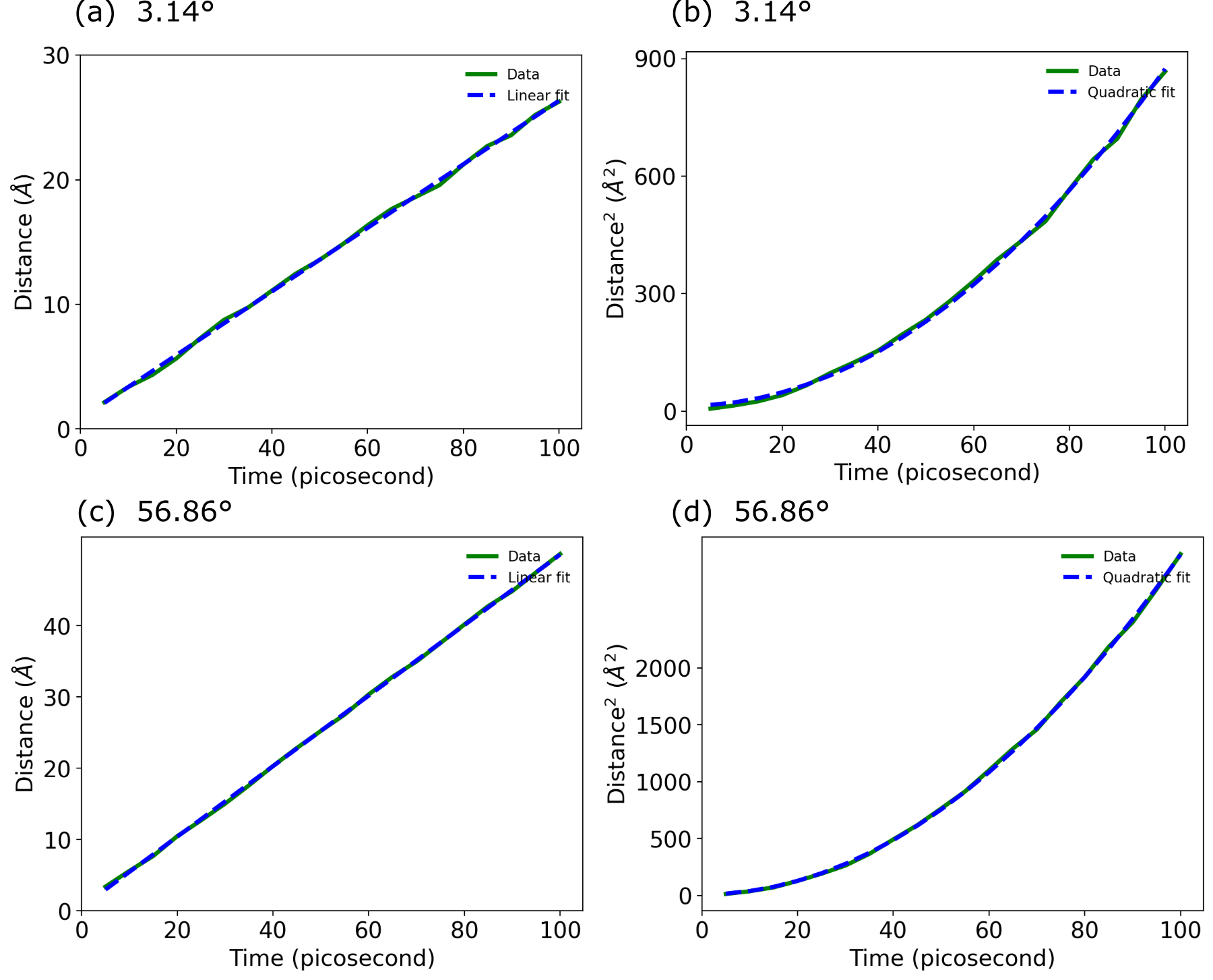}
\caption{Distance and mean-square distance vs. time computed for MoSe$_{2}$/WSe$_{2}$ for 3.14$^\circ$ and 56.86$^\circ$ twist angles.} 
\end{figure}

\clearpage 
\newpage 

\section{VI : Phonon dispersion of $3.14^\circ$ twisted MoSe$_{2}$/WSe$_{2}$}
\begin{figure}[h!]
\centering
\includegraphics[width=0.8\textwidth]{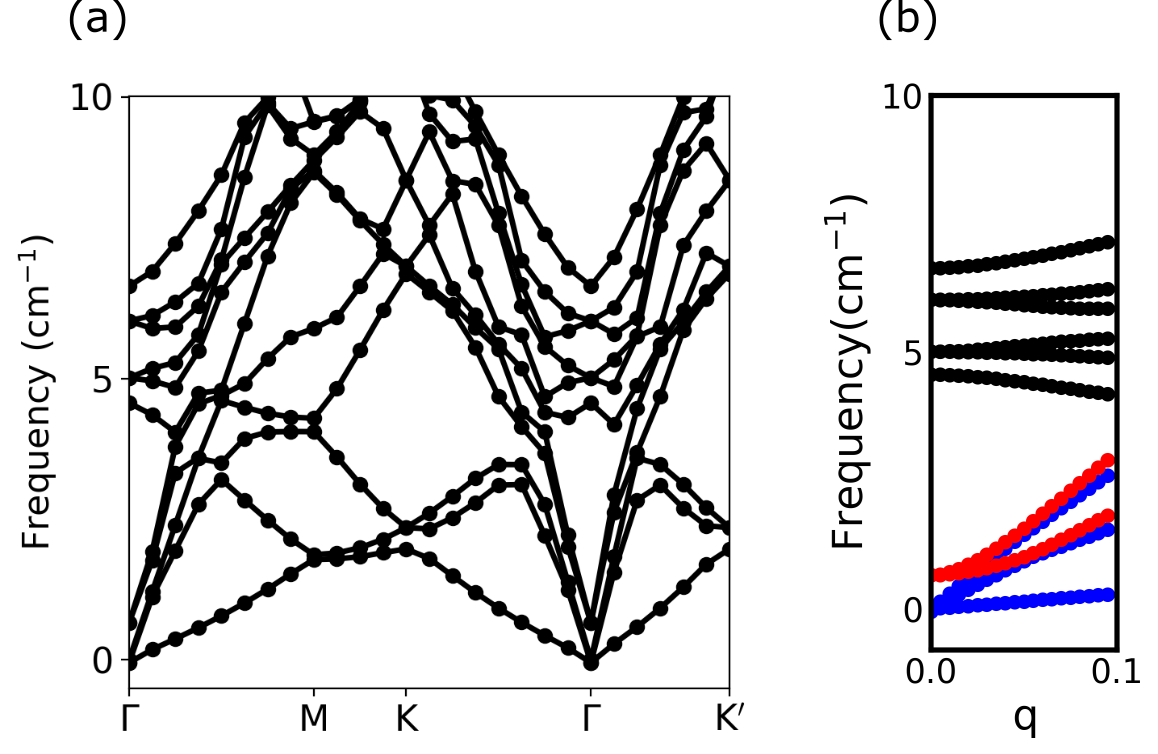}
\caption{(a) Phonon dispersion of the $3.14^\circ$ twisted MoSe$_{2}$/WSe$_{2}$ calculated at T=0 K. (b) Phonon dispersion for very close to the $\Gamma$ point along the $\Gamma-K$ direction (in crystal coordinates). The conventional acoustic modes are highlighted with blue colour. On the other hand, the phason modes are highlighted with red colour. We also extract the group velocity by fitting a line to the linearly dispersing phason modes. The phason velocities are $7.8\times 10^{2}$ mtr./sec. and $3.4 \times 10^{2}$ mtr./sec., respectively. The group velocities associated with the phason modes significantly decrease very close to the $\Gamma$ point as can be seen by deviation from the straight line behaviour. This is in sharp contrast to the acoustic modes (LA/TA modes), where the group velocity remains constant up to arbitrarily small momentum.} 
\end{figure}

\clearpage 
\newpage 

\section{VII: Simulation cell size dependence of the surfing speed for $3.14^\circ$ twisted MoSe$_{2}$/WSe$_{2}$}
\begin{figure}[h!]
\centering
\includegraphics[width=0.6\textwidth]{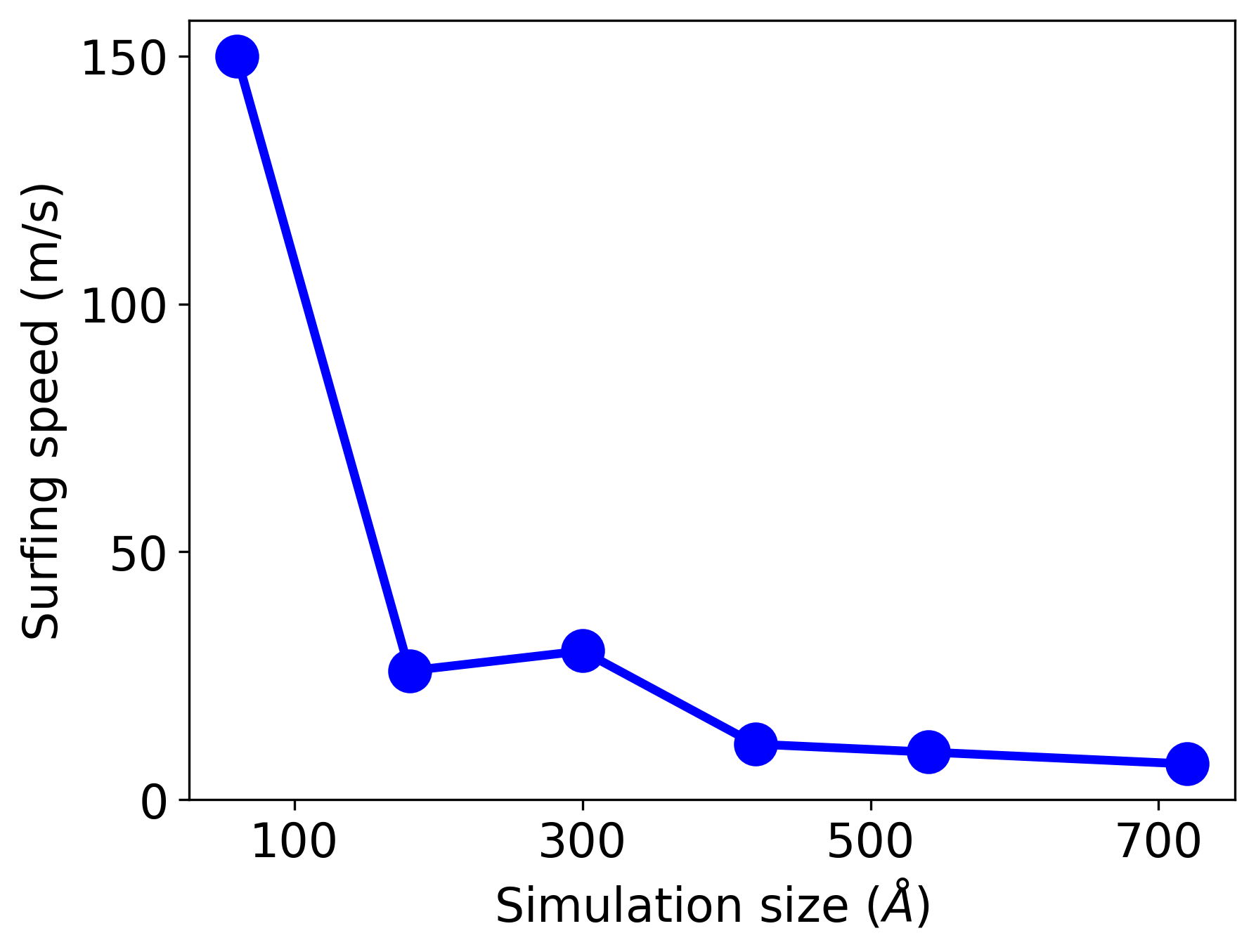}
\caption{The surfing speed is sensitive to the simulation size. However, the surfing persists even for the largest simulation cell used in our calculations, corresponding to a supercell lattice constant of 0.12$\mu m$. } 
\end{figure}

\section{VIII: Surfing in the presence of a quenched disorder at 1200 K}

The attached video (\textit{Surfingindisorder.mp4}) shows the atomic motion (obtained from a molecular dynamics simulation) in the presence of  disorder as described in the main text at a high temperature of 1200~K. We find that surfing survives even in the presence of disorder. Also, we present the same dynamics (\textit{Surfingindisordersideview.mp4}) from a side view focusing on the atomistic details to emphasize the moir\'{e} magnification that happens at the moir\'{e} scale. 


\end{document}